\documentclass[superscriptaddress,showpacs,noshowkeys,twoside,floatfix,reprint]
{revtex4-1}
\usepackage[utf8]{inputenc}
\usepackage{amsmath}
\usepackage{amssymb,epsfig}
\usepackage[dvips]{color}
\usepackage[english]{babel}
\usepackage{isotope}

\begin{document}

  \title{Search for $2p$ decay of the first excited state of \isotope[17]{Ne}}

  \author{P.G.~Sharov}
  \thanks{Corresponding author}
  \email{sharovpavel@jinr.ru}
  \affiliation{Flerov Laboratory of Nuclear Reactions, JINR, Dubna, RU-141980
    Russia}
  \affiliation{SSC RF ITEP of NRC ``Kurchatov Institute'',
    Moscow RU-117218, Russia}
  \author{A.S.~Fomichev}
  \affiliation{Flerov Laboratory of Nuclear Reactions, JINR, Dubna, RU-141980
    Russia}
  \affiliation {Dubna State University, Dubna RU-141982, Russia}
  \author{A.A.~Bezbakh}
  \affiliation{Flerov Laboratory of Nuclear Reactions, JINR, Dubna, RU-141980
    Russia}
  \affiliation{SSC RF ITEP of NRC ``Kurchatov Institute'',
    Moscow RU-117218, Russia}
  \author{V.~Chudoba}
  \affiliation{Flerov Laboratory of Nuclear Reactions, JINR, Dubna, RU-141980
    Russia}
  \affiliation{Institute of Physics, Silesian University in Opava,
    74601 Opava, Czech Republic}
  \author{I.A.~Egorova}
  \affiliation{Bogoliubov Laboratory of Theoretical Physics, JINR, Dubna,
    RU-141980 Russia}
  \affiliation{SSC RF ITEP of NRC ``Kurchatov Institute'',
    Moscow RU-117218, Russia}
  \author{M.S.~Golovkov}
  \affiliation{Flerov Laboratory of Nuclear Reactions, JINR, Dubna, RU-141980
    Russia}
  \affiliation {Dubna State University, Dubna RU-141982, Russia}
  \author{T.A.~Golubkova}
  \affiliation{Advanced Educational and Scientific Center,
    Moscow State University, Kremenchugskaya 11, 121357 Moscow, Russia}
  \affiliation{SSC RF ITEP of NRC ``Kurchatov Institute'',
    Moscow RU-117218, Russia}
  \author{A.V.~Gorshkov}
  \affiliation{Flerov Laboratory of Nuclear Reactions, JINR, Dubna, RU-141980
    Russia}
  \affiliation{SSC RF ITEP of NRC ``Kurchatov Institute'',
    Moscow RU-117218, Russia}
  \author{L.V.~Grigorenko}
  \affiliation{Flerov Laboratory of Nuclear Reactions, JINR, Dubna, RU-141980
    Russia}
  \affiliation{National Research Center ``Kurchatov Institute'',
    Kurchatov sq.~1, RU-123182 Moscow, Russia}
  \affiliation{National Research Nuclear University ``MEPhI'',
    Kashirskoye shosse 31, 115409 Moscow, Russia}
  \author{G.~Kaminski}
  \affiliation{Flerov Laboratory of Nuclear Reactions, JINR, Dubna, RU-141980
    Russia}
  \affiliation{Henryk Niewodniczanski Institute of Nuclear Physics, Polish Academy
    of Sciences, 31342 Cracow, Poland}
  \author{A.G.~Knyazev}
  \affiliation{Flerov Laboratory of Nuclear Reactions, JINR, Dubna, RU-141980
    Russia}
  \affiliation{SSC RF ITEP of NRC ``Kurchatov Institute'',
    Moscow RU-117218, Russia}
  \author{S.A.~Krupko}
  \affiliation{Flerov Laboratory of Nuclear Reactions, JINR, Dubna, RU-141980
    Russia}
  \affiliation{SSC RF ITEP of NRC ``Kurchatov Institute'',
    Moscow RU-117218, Russia}
  \author{M.~Mentel}
  \affiliation{Flerov Laboratory of Nuclear Reactions, JINR, Dubna, RU-141980
    Russia}
  \affiliation{AGH University of Science and Technology,
    Faculty of Physics and Applied Computer Science,
    al.\ Mickiewicza 30, 30-059 Krakow, Poland}
  \author{E.Yu.~Nikolskii}
  \affiliation{National Research Center ``Kurchatov Institute'',
    Kurchatov sq.~1, RU-123182 Moscow, Russia}
  \affiliation{Flerov Laboratory of Nuclear Reactions, JINR, Dubna, RU-141980
    Russia}
  \author{Yu.L.~Parfenova}
  \affiliation{Flerov Laboratory of Nuclear Reactions, JINR, Dubna, RU-141980
    Russia}
  \affiliation{Skobeltsyn Institute of Nuclear Physics, Moscow State University,
    119991 Moscow, Russia}
  \author{P.~Pluchinski}
  \affiliation{Flerov Laboratory of Nuclear Reactions, JINR, Dubna, RU-141980
    Russia}
  \affiliation{AGH University of Science and Technology,
    Faculty of Physics and Applied Computer Science,
    al.\ Mickiewicza 30, 30-059 Krakow, Poland}
  \author{S.A.~Rymzhanova}
  \affiliation{Flerov Laboratory of Nuclear Reactions, JINR, Dubna, RU-141980
    Russia}
  \affiliation{SSC RF ITEP of NRC ``Kurchatov Institute'',
    Moscow RU-117218, Russia}
  \author{S.I.~Sidorchuk}
  \affiliation{Flerov Laboratory of Nuclear Reactions, JINR, Dubna, RU-141980
    Russia}
  \author{R.S.~Slepnev}
  \affiliation{Flerov Laboratory of Nuclear Reactions, JINR, Dubna, RU-141980
    Russia}
  \author{S.V.~Stepantsov}
  \affiliation{Flerov Laboratory of Nuclear Reactions, JINR, Dubna, RU-141980
    Russia}
  \author{G.M.~Ter-Akopian}
  \affiliation{Flerov Laboratory of Nuclear Reactions, JINR, Dubna, RU-141980
    Russia}
  \affiliation {Dubna State University, Dubna RU-141982, Russia}
  \author{R.~Wolski}
  \affiliation{Flerov Laboratory of Nuclear Reactions, JINR, Dubna, RU-141980
    Russia}
  \affiliation{Institute of Nuclear Physics PAN, Radzikowskiego 152, PL-31342
    Krak\'{o}w, Poland}

  \begin{abstract}
    Two-proton decay of the \isotope[17]{Ne} low-lying states populated
    in the $\isotope[1]{H}(\isotope[18]{Ne},d)\isotope[17]{Ne}$
    transfer reaction was studied.
    The two-proton width \(\Gamma_{2p}\) of the \isotope[17]{Ne} first excited
    $3/2^-$ state at $E^*=1.288$ MeV is of importance for the two-proton
    radioactivity theory and nuclear-astrophysics applications.
    A dedicated search for the two-proton emission of this state was performed
    leading to the new upper limit obtained for the width ratio
    $\Gamma_{2p}/\Gamma_{\gamma} < 1.6(3) \times 10^{-4}$.
    A novel, ``combined mass'' method is suggested and tested capable to improve
    the resolution of the experiment
    which is a prime significance for the study of nuclear states with extremely small
    particle-to-gamma width ratios $\Gamma_{\mathrm{part}}/\Gamma_{\gamma}$.
    The condition  $\Gamma_{\mathrm{part}} \ll \Gamma_{\gamma}$
    is quite common for the states of astrophysical interest
    which makes the proposed approach promising in this field.
  \end{abstract}


  \maketitle


\section{Introduction}


One of the most important tasks for nuclear studies in astrophysics
is the measurement of various reaction rates for nucleosynthesis calculations.
In particular, for the important situation of the \emph{resonant}
particle radiative capture the reaction rate of the selected resonant state at
temperature $T$ is proportional to
\begin{equation}
  \langle \sigma_{\mathrm{part},\gamma} \rangle (T) \sim \frac{1}{T^{3n/2}}\,
  \exp \left(-\frac{E_r}{kT} \right)\,
  \frac{\Gamma_{\gamma} \Gamma_{\mathrm{part}}} {\Gamma_{\mathrm{tot}}},
\label{eq:rate}
\end{equation}
where $E_r$ is the resonance position, $\Gamma_{\gamma}$ and
$\Gamma_{\mathrm{part}}$ are partial widths of the resonance $E_r$
into gamma and particle channels \cite{Fowler:1967}.
The ``particle'' here can be proton, alpha, two protons, etc,
and $n$ is the number of captured particles: 1 for $p$, $\alpha$ and 2 for $2p$ captures.
The total width of the resonance in the majority of cases is defined as
$\Gamma_{\mathrm{tot}} = \Gamma_{\gamma} + \Gamma_{\mathrm{part}}$.
Therefore in all such cases there are just two values needed to define
the resonance contribution to the radiative capture rate:
(i) the resonance energy $E_r$ and
(ii) the ratio $\Gamma_{\mathrm{part}} / \Gamma_{\mathrm{tot}}$.
For the resonant states situated well below the Coulomb barrier
particle width can be very small, this ratio then becomes effectively equal to
$\Gamma_{\mathrm{part}} / \Gamma_{\gamma}$,
 and its determination becomes a complicated task.

Reaction studies provide a way to determine the
$\Gamma_{\mathrm{part}} / \Gamma_{\mathrm{tot}}$
ratio without $\gamma$-measurements.
The comparison of data obtained in the measurements made for the missing mass
and invariant mass spectra represents one of the possible ways to obtain the
$\Gamma_\mathrm{part}/\Gamma_\gamma$ ratio.
The missing mass spectrum allows the definition of the population cross section
for the state of interest, while the invariant mass spectrum provides
the decay cross section into the particle channel.
The ratio of these cross sections is evidently equal to
 $\Gamma_{\mathrm{part}}/(\Gamma_{\mathrm{part}}+\Gamma_{\gamma})$.

In this work we address the subject of the extremely weak $2p$ decay branch
of the first excited $3/2^-$ state of \textsuperscript{17}Ne
which is known~\cite{Guimaraes:1995a,Guimaraes:1995b}
to predominantly undergo \(\gamma\) decay to the ground state.
Below in Section~\ref{sec:problem} we discuss some issues of
the \isotope[17]{Ne} study in general and then return to the subject of measuring
the extremely small $\Gamma_{\mathrm{part}} / \Gamma_{\gamma}$ ratio in Section~\ref{sec:approach}.
Further in this work, we study experimentally the
$\isotope[1]{H}(\isotope[18]{Ne},d)\isotope[17]{Ne}$
reaction and derive a new upper limit for the $\Gamma_{2p}/\Gamma_{\gamma}$
ratio of the $3/2^-$ state.
Instead of the \emph{invariant mass} method to determine the $2p$-decay rate
we use the novel \emph{combined mass} method to provide better experimental resolution.
In Section~\ref{sec:prospects} we demonstrate the prospects of
the developed method in forthcoming experiments to reduce this limit to the range where,
according to theoretical predictions,
the direct observation of the $3/2^-$ state $2p$ decay becomes realistic.

\begin{figure}
\begin{center}
\includegraphics[width=0.45\textwidth]{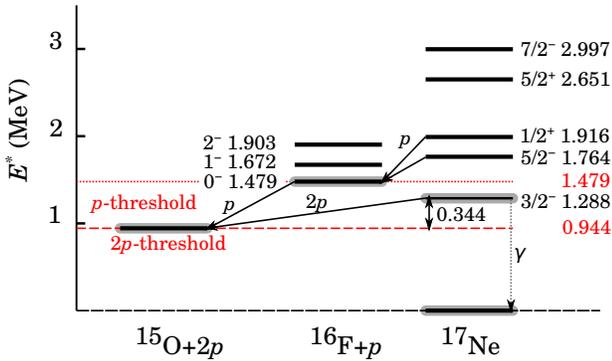}
\end{center}
\caption{\label{fig:levels}
  The level schemes for \isotope[17]{Ne}, its one-proton subsystem \isotope[16]{F},
  and decay scheme for \isotope[17]{Ne} states.}

\end{figure}


\section{Problems of $\isotope[17]{Ne}$ studies}
\label{sec:problem}


Neon-17 is a kind of ``conundrum nucleus''
whose structure and reactions attracted a lot of interest.
Multiple efforts to investigate it, both theoretically and experimentally,
have not yet provided convincing clarity about its properties.
There are several questions of special interest connected with this nucleus
which are actually tightly interwoven.

The \isotope[17]{Ne} nucleus sits on the proton dripline and it is relatively loosely
bound with respect to the \(2p\) breakup
($E_b=944$ keV, see Fig.~\ref{fig:levels}).
This is a borromean nucleus: its \isotope[16]{F} subsystem is unbound
with $S_p(\isotope[16]{F})=-535$ keV.
The situation with \isotope[17]{Ne} separation energies is highly analogous to that with the
``classical'' two-neutron halo nucleus \isotope[6]{He} and thus the question about the
existence of the two-proton halo in \isotope[17]{Ne} was formulated in Ref.~\cite{Zhukov:1995c}.
The intrigue of this problem is enhanced by the fact that
the \isotope[17]{Ne} nucleus is probably the only realistic candidate to possess
a two-proton halo
as in heavier proton dripline systems the specific spatial extension
of the valence nucleon wave functions should be suppressed by the Coulomb interaction.
The issue of two-proton halo in \isotope[17]{Ne} was discussed in
the theory papers\ \cite{Grigorenko:2003,Grigorenko:2005}
and experimental works \cite{Marganiec:2012,Marganiec:2016}.

The \isotope[15]{O} nucleus is a ``waiting point'' in the astrophysical rp-process
as its half-life $T_{1/2}=122.24$ s is comparable to the timescale of
the typical rp-process scenarios.
The radiative absorption of two protons (\(2p\) capture) is known to be
a possible bypath for this waiting point \cite{Gorres:1995}.
The so far unknown \(2p\) decay width of the first excited state of
\isotope[17]{Ne} (\(E^*=1288\) keV, \(J^{\pi}=3/2^-\)) makes a key point
for solving the bypass problem of the 15O waiting point.o this issue.
Taking into account this (previously omitted) state in the calculation of the
\emph{resonant} radiative capture rate strongly modified the corresponding rate
in a broad temperature range around 0.15 GK \cite{Grigorenko:2005a}.
This modification is as large as $3-8$ orders of magnitude,
where the variation corresponds to the uncertainty in $2p$ width predicted in
theory works \cite{Grigorenko:2005a,Grigorenko:2007} the 1.288 MeV $3/2^-$ state.

There is considerable interest in studies of the $2p$ decay of
the $3/2^-$ state from the nuclear theory side as well.
The two-proton decay energy of the first excited state in \isotope[17]{Ne} is only
$E_T=-S_{2p}(\isotope[17]{Ne})= 344$ keV.
This state of \isotope[17]{Ne} belongs to the class of so-called
``true'' two-proton emitters~\cite{Pfutzner:2012}.
Here the protons should be emitted simultaneously because
the narrow ($\Gamma \sim 40$ keV) ground state of
the intermediate \isotope[16]{F} system at $E_r=535$ keV is not accessible for sequential decay.
Due to the very small $2p$ decay energy $E_T$,
the two-proton decay of \isotope[17]{Ne}
should have a typical radioactivity-scale lifetime.
For many years preceding the discovery of the ground state $2p$ radioactivity
in heavier proton dripline nuclei \isotope[45]{Fe}, \isotope[48]{Ni}, and \isotope[54]{Zn}
Ref.~\cite[and Refs.\ therein]{Pfutzner:2012}, the first excited state of
\isotope[17]{Ne} nucleus remained along the prime candidates for
the discovery of $2p$ radioactivity.

Shell-model calculation using the WBP interaction led to an estimated partial decay width of
$5.5 \times 10^{-9}$ MeV for the $\gamma$ decay of \isotope[17]{Ne} $3/2^-$ state \cite{Chromik:1997}.
The \(2p\) decay from the $3/2^-$ state to the $1/2^-$ ground state of
\isotope[15]{O} can proceed only via the escape of the $sd$-shell proton pair.
Few-body calculations performed in the two different theoretical approaches predicts the
two-proton decay width to be $\Gamma_{2p}\sim (5-8) \times 10^{-15}$ MeV
\cite{Grigorenko:2007} or $\Gamma_{2p}=1.4 \times 10^{-14}$ MeV \cite{Garrido:2008}.
For
the first excited state of \isotope[17]{Ne} this gives a ratio of
$\Gamma_{2p}/\Gamma_{\mathrm{tot}} \approx \Gamma_{2p}/\Gamma_{\gamma}$ to be $(0.9-2.5)\times 10^{-6}$.

Up to now only an upper limit of this value is known \cite{Chromik:2002}.
From the non-observation of the two-proton emission from the $3/2^-$ state
a one-sigma upper limit for
the branching ratio $\Gamma_{2p}/\Gamma_{\gamma} \le 7.7\times 10^{-3}$ was set.


\section{Experimental approach}
\label{sec:approach}


The searched $\Gamma_{2p}/\Gamma_{\gamma}$ branching ratio for
the $2p$ decay of the \isotope[17]{Ne} $3/2^-$ state is expected to be located in a broad band of
values from the experimental limit
$\Gamma_{2p}/\Gamma_{\gamma} < 7.7\times 10^{-3}$
to the theoretical predictions
$\Gamma_{2p}/\Gamma_{\gamma} \sim 9\times 10^{-7}$.
The primary purpose of this work was to reduce the experimental limit
as much as possible and to test the suggested combined mass method.

The one-neutron transfer reaction
\(\isotope[1]{H}(\isotope[18]{Ne},d)\isotope[17]{Ne}\)
is the subject of study in this work.
The measurement of the missing mass spectrum provides the \emph{population rates} for the \isotope[17]{Ne} states.
Standard methodology implies that,
turning to the measurement of the invariant mass spectrum,
one can detect the yield of a weak \(2p\)-decay branch
inherent to the state of interest which is the \(3/2^-\) state of \isotope[17]{Ne}.
Naturally, the revelation of the tiny particle-decay branch of the lower-lying \(3/2^-\) state
requires special attention to the reduction of the background coming
from the strong particle-decay branch of the higher-lying \(5/2^-\) or/and \(1/2^+\) states.

There are two ways to overcome this problem.
(i) Choose the reaction which provides the highest ratio for the population of
the state of interest with respect to the nearest states which can seed
background events in the energy range of interest.
(ii) Maximal increase of the experimental resolution,
so that the background events connected with the nearest states
are well separated from the energy range of interest.

Deciding in favor of the
\(\isotope[1]{H}(\isotope[18]{Ne},d)\isotope[17]{Ne}\) reaction
one can benefit from the fact that the nearest state to $3/2^-$,
the $5/2^-$ state, is expected to be poorly populated
in this reaction.
DWBA calculations (see Section~\ref{sec:data-analysis} for the parameters)
show the average cross section for the $5/2^-$ state,
taken in a proper range center-of-mass angle,
being more than one order of magnitude
less then that for the $3/2^-$ state.

\begin{figure}
\begin{center}
\includegraphics[width=0.42\textwidth]{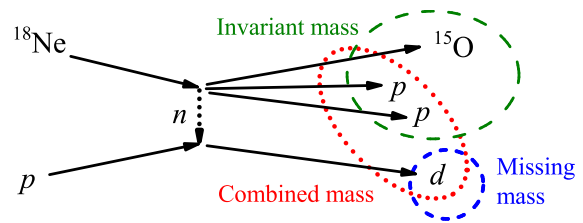}
\end{center}
\caption{\label{fig:combined} (color online)
  The illustration of missing mass,
  invariant mass, and combined mass methods for the
  \(\isotope[1]{H}(\isotope[18]{Ne},d)\isotope[17]{Ne}\) reaction.}
\end{figure}

To enable the better selection of the desired \isotope[17]{Ne} state
we used a novel approach which was named the combined mass method.
The measurement of the emission angle and energy of recoil deuteron appearing
in this reaction is the prerequisite to the realization of this approach.
Figure~\ref{fig:combined} is destined to illustrate, on the example of
the \(\isotope[1]{H}(\isotope[18]{Ne},d)\isotope[17]{Ne}\) reaction,
the choice of particles to be detected in the combined mass method.
Of course, this is the only
objective when the excitation spectrum of \isotope[17]{Ne} is measured by the missing mass method.
The yields of different resonance states are defined by this method irrespective of their decay modes.
The detection of the recoil deuteron in coincidence with protons only offers
a way to the yield determination made for the decay branches of
the \isotope[17]{Ne} excited states associated with the proton emission.
Certainly, of particular interest here is the search for the anticipated
very weak two-proton decay branch of its first exited state
(\(E^*=1288\) keV, \(J^{\pi}=3/2^-\)).

In the proposed approach the ability to select rare $2p$-decay events of
the $3/2^-$ state in \isotope[17]{Ne} is improved due to the
considerable mass asymmetry for the particles both in the reaction
($A_d/A_{\isotope[17]{Ne}}=2/17$) and in the decay ($A_{2p}/A_{\isotope[15]{O}}=2/15$) channels.
In the $d-\isotope[17]{Ne}$ center-of-mass system the velocity of the
heavier particle ($\isotope[17]{Ne}^*$) is known with the precision better by
a factor of $17/2$ than that for the lighter one (deuteron).
The energies of the emitted protons in the
\isotope[17]{Ne} center-of-mass are defined with much better accuracy.
So, having measured in laboratory system the emission angle of the recoil deuteron measured even with a modest
(about one degree) precision one specifies with a tremendous accuracy
(within $\Delta E/E\approx 6\times10^{-4}$ and better than 0.1 degree, respectively)
the energy and escape direction of \isotope[17]{Ne}.
It should be noted that such a situation is typical in general for the study of
transfer reactions made in inverse kinematics with heavy projectiles bombarding
light target nuclei.
The resolution attainable by means of combined mass method for
the excitation energy spectrum depends on the target thickness which can be set
to be quite large due to the small specific energy loss of the protons which are
due to detection.
This favors the revelation of such small \(2p\) decay branch
anticipated for the 1288 keV excited state of \isotope[17]{Ne}.

\section{Experimental setup}


The experiment was performed in the Flerov Laboratory of Nuclear Reactions at JINR.
The \isotope[18]{Ne} beam with energy {35} MeV/nucleon was produced in
fragmentation of a \(53A\) MeV primary beam of \isotope[20]{Ne}
bombarding a 55.5 mg/cm\textsuperscript{2} \isotope[9]{Be} target.
The secondary beam was selected using the ACCULINNA fragment separator \cite{Rodin:1997}.

The standard set of beam diagnostic detectors included two plastic scintillators
(for time-of-flight -- ToF and energy loss -- $\Delta E$ measurements)
and two position-sensitive multiwire chambers
(for beam tracking).
Data presented in this paper were collected in experiments carried out with
a beam of total intensity on target of \(2\times10^5\) pps.
The part of \isotope[18]{Ne} ions in the beam cocktail was about $18\%$.

The secondary beam was focused on a cryogenic hydrogen target.
The target was a cylindrical cell having the two 6 \(\mu\)m stainless-steel
windows entrance and exit.
Two versions of the target cell with different thicknesses were used in this experiment.
The ``thin'' target, intended for work with hydrogen in the gaseous phase,
had windows of 20 mm in diameter and a distance of 4 mm between them.
The gas pressure in the target was 2 bar and
the target was cooled to 24 K.
The ``thick'' target, designed for operation with hydrogen in the liquid phase,
had windows of 30 mm diameter and the effective thickness of $1.1$ mm.

\begin{figure}
\begin{center}
\includegraphics[width=0.48\textwidth]{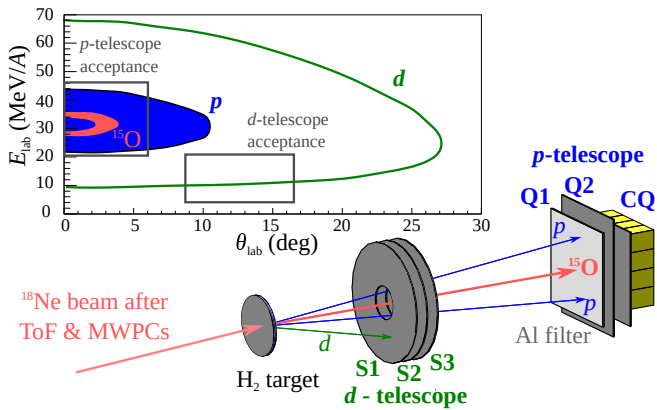}
\end{center}
\caption{(color online) The experimental setup and kinematic plot for the reaction products.
  Explanations are given in text.}
\label{fig:setup}
\end{figure}

Figure~\ref{fig:setup} shows a schematic drawing of the detector setup.
The annular telescope, located at 15 cm downstream the target (``thin'' target case)
and 12 cm downstream the target (``thick'' target case),
detected the recoil deuterons from the \(\isotope[1]{H}(\isotope[18]{Ne},d)\) reaction.
The telescope consisted of three position sensitive Si detectors with
an inner (outer) radius of the sensitive area of 16 (41)  mm and a thickness of 1 mm each.
The first double-side silicon detector (S1) was segmented into 16 concentric rings on one side
and 16 sectors on the other side.
The two subsequent silicon strip detectors (S2 and S3), segmented into 16 sectors each,
provided the measurement of total energy.
Particle identification was performed by standard $\Delta E-E$ analysis.
Each of the detector segments had its independent spectrometric channel.
Signals from any sector of the S1 detector triggered the data acquisition system.
An additional trigger (signal from the second ToF scintillator selected with
a counting rate reduction of 4096) was used for beam monitoring.

Another telescope located on the beam axis at a distance of 30 cm from the target
was intended for the detection of protons from the
$\isotope[17]{Ne}^*\to \isotope[15]{O}+2p$ decay.
The telescope consisted of two square $6\times 6$ cm$^{2}$, 1 mm thick silicon detectors (Q1 and Q2)
segmented into 32 strips on one side.
Following the pair of Si detectors installed was a wall of 16 CsI(Tl) crystals with PMT (Hamamatsu R9880U-20) readout (CQ).
Each crystal was $1.6 \times 1.6$ cm$^{2}$ across and had a thickness of 3.0 cm.
To ensure the normal working conditions for the detectors
a 1.4 mm thick aluminum filter was installed directly in front of the telescope.
This was enough to stop all the nuclei making the beam cocktail while the protons
from the decay of $\isotope[17]{Ne}^*$ lost only a small part of their energy in the aluminum filter.


\section{Data analysis}
\label{sec:data-analysis}

\begin{figure}
\begin{center}
\includegraphics[width=0.48\textwidth]{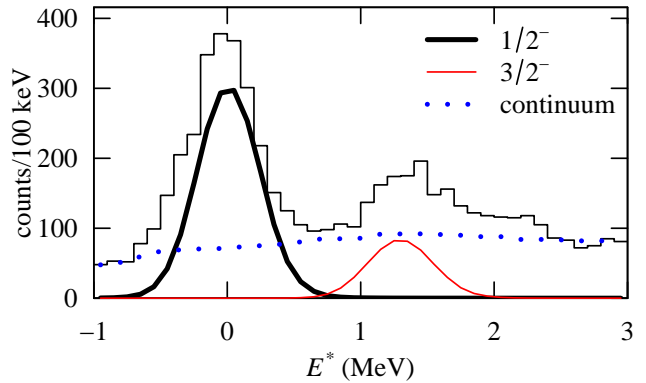}
\end{center}
\caption{Missing mass spectrum (histogram)
  of \(\isotope[17]{Ne}^*\)
  measured using the thin gas target in the angular range $4^\circ-18^\circ$ in c.m.s.
  Lines show the calculated yields of the \(1/2^-\) and \(3/2^-\) states
  of \isotope[17]{Ne} and the estimated continuum background.}
\label{fig:thin_target}
\end{figure}

The first part of this experiment was made with thin (gaseous) target.
A total flux of $7.4\times 10^9$ bombarding \isotope[18]{Ne} nuclei
passed through the hydrogen target with thickness $4.8 \times 10^{20}$ cm\textsuperscript{-2}.
Figure~\ref{fig:thin_target} shows the missing mass spectrum of
\isotope[17]{Ne} obtained as a result of this study of
\(\isotope[1]{H}(\isotope[18]{Ne},d)\isotope[17]{Ne}\) reaction.
One can see in this spectrum the separate peak corresponding to the \isotope[17]{Ne} ground state
and the bump with its left side centered close to the position of the first
\((J^{\pi} = 3/2^-)\) excited state.
These structures are superimposed on the smooth continuum
appearing mainly due to the reactions on the target windows.
The bump is centered close to the position of the first excited state
\((J^\pi=3/2^-)\) which points on the low population
of the nearest \(5/2^-\) and \(1/2^+\) states.
Therefore one can  sorts out only the peak corresponding to
the contribution of \(3/2^-\) state to the measured spectrum.

The measured missing mass spectrum allowed us to make a good estimation of
the ground-state yield, \(N=1635(52)\).
Around one third of that value makes the number of events responsible for
the peak attributed to the \(3/2^-\) state population.
To reliably estimate the yield ratio of the two states
(the \(J^\pi=1/2^-\) g.s.\ and the \(J^\pi=3/2^-\) first excited state)
we carried out the two-step DWBA calculations of \(\isotope[18]{O}(d,\isotope[3]{He})\) and
\(\isotope[18]{Ne}(p,d)\) proton (neutron) pick-up reactions.
The existing experimental data on the angular dependence of
the \(\isotope[18]{O}(d,\isotope[3]{He})\) reaction differential cross sections
at 52~MeV~\cite{Hartwig:1971} have been reanalyzed and, using a common exit
\isotope[3]{He} OM potential for these states, the ratio of
Spectroscopic Factor (SFs) of the first excited state of \isotope[17]{N}
to that of the ground state was obtained.
In the case of the \(\isotope[1]{H}(\isotope[18]{Ne},d)\isotope[17]{Ne}\) reaction
studied in the present work we consider only the reaction channels
populating the \(1/2^-\) g.s.\ and the \(3/2^-\) excited state.
In a way similar to the isobaric-analogue case, the cross sections for the
\(\isotope[18]{Ne}(p,d)\) reaction were calculated.
The OM parameters for the reaction entrance channel were taken from
the CH89 global nucleon-nucleus potential algorithm.
For the exit channels the CH89 potentials for proton and neutron were folded
to obtain the deuteron OM potential for the energy corresponding to the
\isotope[17]{Ne} g.s.
The same potential was applied to the \isotope[17]{Ne} first excited state.
The resulting differential cross section values were multiplied by
the corresponding factors emerging from the analysis of data known for the
\(\isotope[18]{O}(d,\isotope[3]{He})\) reaction.

\begin{figure}[htb]
  \centering
  \includegraphics[width=0.48\textwidth]{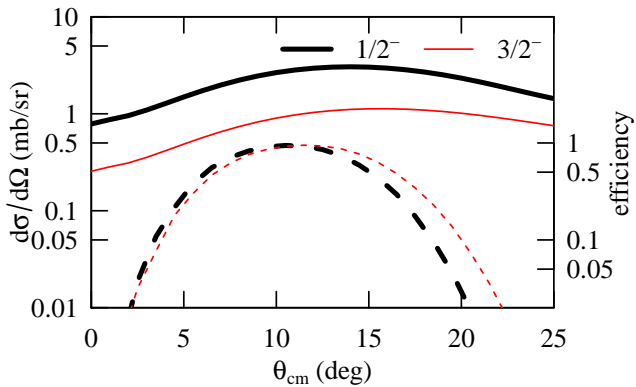}
  \caption{\label{fig:cross-sec} (color online)
    Calculated differential cross sections in c.m.s.\ for the
    \isotope[17]{Ne} states obtained in the
    $\isotope[1]{H}(\isotope[18]{Ne},d)\isotope[17]{Ne}$ reaction (solid lines)
    and the corresponding detection efficiencies provided for deuterons
    emitted in the individual reaction channels (dashed lines).}
\end{figure}

Figure~\ref{fig:cross-sec} shows the calculated differential cross sections
in the center-of-mass system for the $1/2^-$ and $3/2^-$ states of \isotope[17]{Ne}.
The detection efficiencies for the population of these states were obtained
by the Monte-Carlo simulation.
The corresponding curves are shown in Fig.~\ref{fig:cross-sec} by dashed lines.

The yield obtained in our experiment for the \isotope[17]{Ne} ground state
agrees within $\pm 10\%$ with that coming out from
the differential cross section calculated with a value of 1.2 assumed for the spectroscopic amplitude
(lines in Fig.~\ref{fig:thin_target} show the calculated yields of these states,
and the calculated differential cross sections are shown in Fig.~\ref{fig:cross-sec}).
The significant result for our further analysis is the conclusion that one can rely on
the cross section values given in Fig.~\ref{fig:cross-sec} for the reaction
\(\isotope[1]{H}(\isotope[18]{Ne},d)\isotope[17]{Ne}(J^\pi=3/2^-)\).

\begin{figure}[t]
  \begin{center}
    \includegraphics[width=0.48\textwidth]{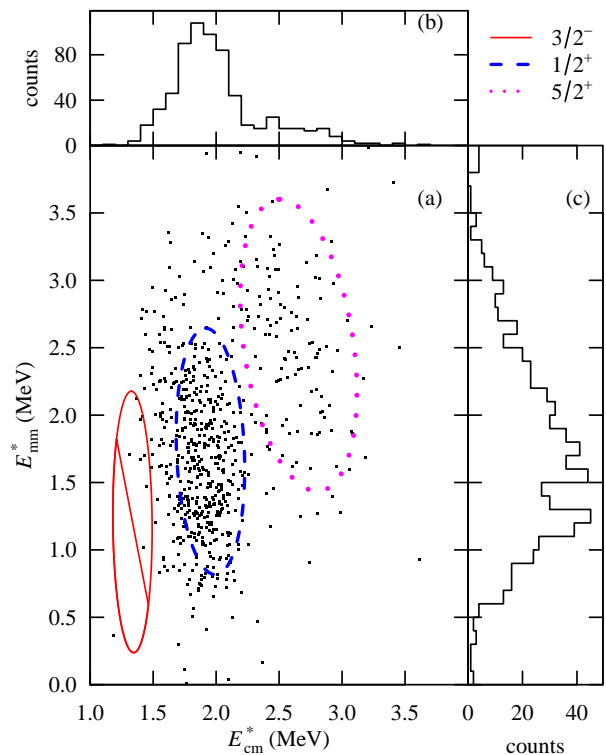}
  \end{center}
  \caption{\label{fig:thk-2p-hits} (color online)
    Excitation energy spectrum of \isotope[17]{Ne}:
    (a) Correlation plot showing the excitation energy of \isotope[17]{Ne}
    measured by the missing mass method (\(E^*_{\mathrm{mm}}\))
    versus the energy obtained by the combined mass method (\(E^*_{\mathrm{cm}}\)).
    The ellipses shown by solid red, dashed blue, and dotted magenta curves
    correspond to the loci where the observation of $68\%$ of events
    for the $3/2^-$, $1/2^+$, and $5/2^+$ states, respectively, is expected.
    (b) combined mass spectrum,
    (c) missing mass spectrum.}
\end{figure}

When analyzing the data collected in the thick-target experiment we took into account
the \(d–p–p\) coincidence events detected under the condition that the \isotope[18]{Ne}
beam nuclei hit the central part of the hydrogen target within a circle with diameter of 18 mm.
This choice eliminated the background caused by protons scattered from the material
at the edges of the target windows.
The so selected beam flux made $2\times 10^{10}$ \isotope[18]{Ne} nuclei
passing through the hydrogen target with thickness \(4.6(4)\times 10^{21}\) cm$^2$.
In total, 660 $d–p–p$ coincidence events were detected in this experiment.

Analysis took into account the energy and the trajectories (angles) of
\isotope[18]{Ne} projectiles on the target.
These parameters were measured with accuracy 1.5\% and 2 mrad, respectively.
It was supposed that the origin of all observed events was in the middle plain of the target.
The triple coincidence events were detected in the center-of-mass angular range $3^\circ – 24^\circ$,
and their detection probability in small angular bins was estimated numerically.
The values of the missing mass and combined mass energies, $E^*_\mathrm{mm}$ and
$E^*_\mathrm{cm}$, emerging from the data of each event were found.

The obtained spectra are presented in Fig.~\ref{fig:thk-2p-hits}.
In panel~\ref{fig:thk-2p-hits}~(a) the 660 events are displayed
in their positions in the $E^*_\mathrm{mm}$ vs $E^*_\mathrm{cm}$ plot.
Projections of this two-dimensional plot on its axes are the combined mass
and missing mass spectra which are presented in panels~\ref{fig:thk-2p-hits}~(b,c).
Poor resolution obtained in the missing mass spectrum emerges
from the distortion caused by the thick target.

The combined mass for each individual $d – p – p$ coincidence event is defined as
the \isotope[17]{Ne} decay energy obtained as the sum of
the center-of-mass energies of the two emitted protons.
The small correction for the \isotope[15]{O} recoil energy is taken into account.
In addition to the energies and emission angles of the two protons
measured well in lab system the analysis requires a good knowledge of
the momentum vector of \isotope[17]{Ne} produced in the
$\isotope[1]{H}(\isotope[18]{Ne},d)\isotope[17]{Ne}$ reaction.
Of key importance here is the emission direction (polar angle)
of the recoil deuteron measured in the experiment with 8 mrad precision.
This defines the energy and polar angle of \isotope[17]{Ne}
with accuracy 0.03\% and 2 mrad, respectively.

Following procedures were implemented in the data analysis.
The excitation energy of \isotope[17]{Ne} is separated in two parts:
the energy of relative motion of two protons ($E_x$) and
the relative-motion energy of the core and di-proton ($E_y$).
The value $E_x$ for each individual event is defined from
the proton momentum vectors measured in the lab system.
Switching to the \isotope[17]{Ne} center-of-mass frame
the \isotope[15]{O} momentum is also found according to momentum conservation,
and $E_y$ emerges from the defined \isotope[15]{O} momentum.
The accuracy of the derived $E_x$ and $E_y$ values is limited
by errors occurring at the transition from the lab system to the center-of-mass system.

The combined mass energy $E_\mathrm{cm} = E_x + E_y$
has accuracy better than the missing mass.
This makes perceptible the two states of \isotope[17]{Ne}
(the states with $J^\pi=1/2^+$, $E^*=1288$ keV and $J^\pi=5/2^+$, $E^*=2651$ keV)
situated above its $2p$-decay threshold (see in Fig.~\ref{fig:thk-2p-hits}(a,b)).
We can not exclude that a few tens of events connected with the \isotope [17]{Ne} $5/2^-$ state,
are present in the spectrum in Fig.~\ref{fig:thk-2p-hits}
but these are indistinguishable among the overwhelming number of the \(1/2^+\) events and
don't affect the further analysis.
The event localization obtained in this experiment in the two-dimensional plot
$E_\mathrm{mm}$ vs $E_\mathrm{cm}$ was tested by Monte-Carlo (MC) simulations
incorporating the standard Geant4 \cite{Agostinelli:2003,Allison:2006}
electromagnetic physics set for the description of particle interaction with matter.
The simulation resulted in the definition of the loci
where the majority of events (68\%) associated with the \isotope[17]{Ne}
excited states are localized.
One can see in Fig.~\ref{fig:thk-2p-hits}~(a) that the event pattern
obtained in experiment is described rather well by the simulation.
It describes also the event seeding towards
the low-energy tails of the $E_\mathrm{cm}$ spectra.
The origin of this seeding effect in the discussed experiment was made by
the small fraction of protons scattered from the annular telescope intended for
the deuteron detection (see this in Fig.~\ref{fig:setup}).

\section{Width ratio evaluation}


The \(\Gamma_{2p}/\Gamma_\gamma\) value can be evaluated by the following equation
\begin{equation}
  \frac{\Gamma_{2p}}{\Gamma_\gamma}= \dfrac{N_{2p}}{\varepsilon_{2p} N} \,,
  \label{eq:g-ratio}
\end{equation}
where \(N_{2p}\) is the number of measured $d$-$p$-$p$ events;
\(N\) --- total number of events where \isotope[17]{Ne} was produced
in its first excited \(3/2^-\) state;
\(\varepsilon_{2p}\) --- detection efficiency of the two protons
emitted by \isotope[17]{Ne}.
The value of \(\varepsilon_{2p}\) was estimated by means of Monte-Carlo simulation.
The \(N\) value was estimated from the cross section shown on Fig.~\ref{fig:cross-sec}
\[
  N=N_\mathrm{B}N_\mathrm{at}
  \int\varepsilon \dfrac{d\sigma}{d\Omega} d\Omega =38(6)\times 10^3,
\]
where \(N_\mathrm{B}\) is the number of \isotope[18]{Ne}
beam nuclei hitting the target in this experiment;
\(N_\mathrm{at}\)
is the effective thickness of liquid hydrogen target;
\(\varepsilon\) --- efficiency of deuteron detection
obtained from the Monte-Carlo simulation of the setup.

Main problem of \(N_{2p}\) estimation is that there is no peak associated with
the \(3/2^-\) state in the \(2p\)-coincidence spectrum, see Fig.~\ref{fig:thk-2p-hits}~(b).
The spectrum shows some events near the energy of the \(3/2^-\) state.
Events located in the \(3/2^-\) state energy range can be connected
with this state as well as with the decay of the higher-lying excited states, and
there is no way to clearly separate them.
Therefore we can only set an upper limit for the \(\Gamma_{2p}/\Gamma_\gamma\) ratio of the \(3/2^-\) state.

The excitation energies obtained in the combined mass method (\(E^*_\mathrm{cm}\)) and
missing mass method (\(E^*_\mathrm{mm}\)) should show some correlation.
In particular, events occurring at excitation energy $E^*_\mathrm{mm}>2.2$ MeV
in the spectrum of Fig.~\ref{fig:thk-2p-hits}~(c) have nothing to do with
the $3/2^-$ state population.
Therefore it is appropriate to inspect the locus of this state in
the two-dimensional plot in Fig.~\ref{fig:thk-2p-hits}~(a)
shown in coordinates \(E^*_\mathrm{cm}\) vs \(E^*_\mathrm{mm}\).
The solid ellipse shows the locus, where $68\%$ of \(3/2^-\) state are concentrated.
One can see eight events in this locus, that corresponds to
\(\Gamma_{2p}/\Gamma_\gamma<6(1)\times 10^{-4}\).
However, the majority of events are located in
the top-right part of the ellipse.
If we reduce the \(3/2^-\) locus to bottom-left half of the ellipse delimited by
line $E^*_\mathrm{mm}= 2.5-1.5 E^*_\mathrm{cm}$ then half of
the expected $3/2^-$ events should to be lost,
but there are no events in this region at all.
This gives a lower limit \(\Gamma_{2p}/\Gamma_\gamma<1.6(3)\times 10^{-4}\).


\section{Prospects to improve sensitivity limit of the method}
\label{sec:prospects}


As we can see the $2p$-decay width limit achieved for the first excited state of
\isotope[17]{Ne} is still much higher than
the up-to-date theoretical predictions \cite{Grigorenko:2007,Garrido:2008} give.
It is natural to address a question:
which $2p$-width limit is attainable
just by a realistic optimization of the setup.
We assume here that a \isotope[18]{Ne} beam intensity of \(10^6\) s$^{-1}$
hitting the hydrogen target will be available, and
one should optimize only the energy resolution to get rid of the background
coming from the higher-lying excited states.

\begin{figure}[t]
\begin{center}
\includegraphics[width=0.45\textwidth]{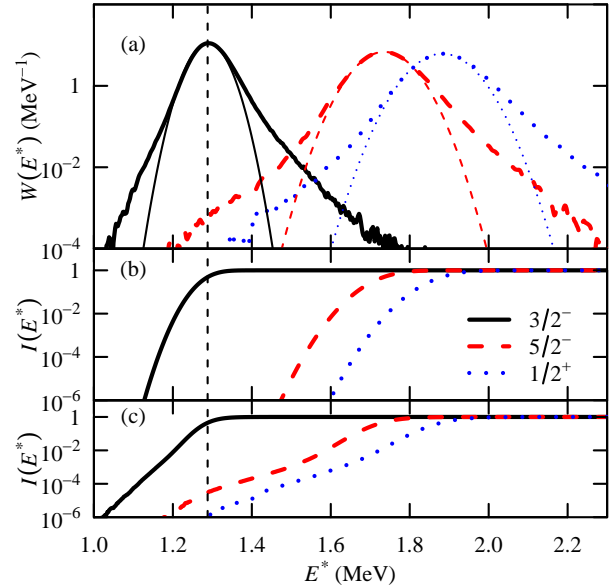}
\end{center}
\caption{Monte-Carlo simulations of the improved setup demonstrating potential sensitivity limit of the method.
The $3/2^-$, $5/2^-$, and $1/2^+$ contributions are shown by solid black,
dashed red, and dotted blue curves, respectively.
Panel (a) shows the probability density $W$.
Thin curves give the Gaussian fits with respective deduced FWHM values of 80, 130, and 140 keV.
Panels (b) and (c) show the cumulative distribution functions $I$
for the Gaussian fits and for the complete MC distributions, respectively.}
\label{fig:limit}
\end{figure}

Figure~\ref{fig:limit} shows the Monte-Carlo simulations of the ``improved'' setup.
The prime parameters of the setup which affect the energy resolution are target thickness, energy and angular resolutions of the charged particle telescopes.
The optimized setup has gaseous hydrogen target with thickness of 1.2 mg/cm\textsuperscript{2}.
Supposition was made that the used recoil-deuteron detector array ensured measurements
for the $\isotope[1]{H}(\isotope[18]{Ne},d)\isotope[17]{Ne}$ reaction taking place
in an angular range of $8^\circ\le \theta_{cm} \le 24^\circ$.

Standard Geant4 \cite{Agostinelli:2003,Allison:2006} electromagnetic physics set was used for particle interaction with matter description.
Panel (a) shows the Monte-Carlo simulation results normalized to the unity
probability density $W(E^*)$ for the $3/2^-$, $5/2^-$, and $1/2^+$ states.
The respective energy resolutions are 80, 130, and 140 keV (FWHM).
Fig.~\ref{fig:limit} (b) and (c) show the cumulative distribution functions
\[
I(E^*) = \int_{-\infty}^{E^*} W(E) \, dE \,,
\]
for Gaussian fits of Fig.~\ref{fig:limit} (a) and for full MC results, respectively.

If we look at the FWHM values only, the achieved resolutions are more than sufficient for our task: the states are very far from overlap, which is confirmed by Fig.~\ref{fig:limit} (b).
But it can be seen in Fig.~\ref{fig:limit}
(a) that the behavior of the distribution ``wings'' deviates from
the Gaussian profile at certain level and extends much further,
beyond the actual resonance position than expected from the resonance FWHM.
It can be found in Fig.~\ref{fig:limit}
(c) that the cumulative contribution of the ``background'' events from the $5/2^-$ and $1/2^+$ states is substantial in the energy region of the $3/2^-$ resonance, achieving $10^{-4}-10^{-5}$.
Fig.~\ref{fig:limit} (c) gives a guideline how to optimize the ``signal to noise'' ratio for the measurements aiming at the $3/2^-$ state decay: we should make cutoff for $3/2^-$ events below certain energy $E^*$.
If we recall that the population of the $5/2^-$ state is at least one order
of magnitude smaller than that for the $3/2^-$ state in the $(p,d)$ reaction,
an optimistic background condition  $4 \times 10^{-6}$ is obtained.
This is quite close to the theoretical limit $\Gamma_{2p}/\Gamma_{\gamma}\sim (0.9-2.5) \times 10^{-6}$ showing that direct $2p$ measurements could be feasible at least from the background point of view.

It is clear that precise behavior of the ``non-Gaussian'' component of the energy resolution is a very delicate issue depending on the fine details of the experimental apparatus.
Here we demonstrate that (i) this issue can be overcome in realistic scenario and the further search for the $2p$ branch in \isotope[17]{Ne} $3/2^-$ state is feasible and (ii) the prior careful studies of this aspect of the planned experiment are unavoidable.

We should also point out that additional background-elimination method can further improve the situation here.
This is connected with energy correlation analysis as provided by Fig.~\ref{fig:thk-2p-hits}(a).
A six-fold improvement was achieved for the $\Gamma_{2p}/\Gamma_{\gamma}$ limit in the current experiment,
and it can be evaluated that even higher improvement factors are possible
depending on the details of the experimental setup.

From a more general point of view, in the particular case of our experiment we aimed to determine the $\Gamma_{\mathrm{part}}/\Gamma_{\mathrm{tot}}$ ratio for the quite complicated $2p$ decay process.
It is clear that in the easier cases of states decaying via proton or
alpha emission the $\Gamma_{\mathrm{part}}/\Gamma_{\mathrm{tot}}$ ratio attainable in this approach should be much better.


\section{Conclusions}


We performed a dedicated search for the $2p$ decay branch of the first excited
$3/2^-$ state of \isotope[17]{Ne} populated in the
$\isotope[1]{H}(\isotope[18]{Ne},d)\isotope[17]{Ne}$ transfer reaction.
The population of low-energy states (\(E^* < 3\) MeV)
in the \isotope[17]{Ne} nucleus was also studied.
Based on these results the new upper limit
$\Gamma_{2p}/\Gamma_{\gamma} \leq 1.6(3) \times 10^{-4}$ is established.
This significantly (about a factor of 50) reduces the value of the limit defined
in the previous work \cite{Chromik:2002}.
The strong improvement of the $\Gamma_{2p}/\Gamma_{\gamma}$ limit was achieved due to the choice of the transfer reaction used
as a tool for the two-proton decay study of \(\isotope[17]{Ne}^*\) and
application of the novel ``combined mass'' method to the reconstruction of
the \isotope[17]{Ne} excitation spectrum.
The latter allowed us to improve significantly the instrumental resolution in
the measurements made with the thick target.
The measured limit for the rate value rules out the predictions made for
the $2p$ decay width of the \isotope[17]{Ne} first excited state
by the simplified di-proton decay model \cite{Chromik:1997}, but it is still insufficient to be restrictive
for the realistic theoretical predictions \cite{Grigorenko:2007,Garrido:2008}.

We see prospects for a considerable (by 1-2 orders of magnitude) reduction of
the $\Gamma_{2p}/\Gamma_{\gamma}$ upper limit in the proposed experimental method
without revolutionary modification of the setup.
Such improvements open a way to the direct experimental observation of the true,
radioactive $2p$-decay of the \isotope[17]{Ne} $3/2^-$ state taking
the theoretically predicted ratio of
$\Gamma_{2p}/\Gamma_{\gamma} \sim (0.9-2.5) \times 10^{-6}$ as a trusted aim.

The issue of general interest is the development of methods applicable to
the studies of weak particle (alpha, proton, or two-proton) decay branches
of excited states which reside well below the Coulomb barrier and thus have
extreme small $\Gamma_{\mathrm{part}}/\Gamma_{\gamma}$ ratios.
The possibility to derive directly such weak decay branches in one experiment makes
promising the application of the proposed approach to the problems of nuclear astrophysics.


\begin{acknowledgments}


A.A.B., S.A.R., T.A.G., A.V.G., I.A.E., S.A.K., A.G.K., and P.G.S.
are supported by the Helmholtz Association under grant agreement IK-RU-002.
This experiment was partly supported by the
RSF 17-12-01367 grant and MEYS Czech Republic LTT17003 and LM2015049 grants.
The authors are grateful to Profs. Yu.Ts. Oganessian and S.N. Dmitriev for
support to this work,
helpful discussions with Dr. I. Mukha are acknowledged as well.
\end{acknowledgments}


\bibliography{ref}


\end{document}